# Spontaneous order in ensembles of rotating magnetic droplets

A. P. Stikuts[a,b,*], R. Perzynski[b] and A. Cēbers[a,c]

[a]MMML Lab, Faculty of Physics and Mathematics, University of Latvia, Jelgavas iela 3 - 014, Riga, LV-1004, Latvia
[b]Sorbonne University, CNRS, UMR 8234, PHENIX, Paris, F-75005, France
[c]Chair of Theoretical Physics, University of Latvia, Jelgavas iela 3, Riga, LV-1004, Latvia

ABSTRACT

Ensembles of elongated magnetic droplets in a rotating field are studied experimentally. In a given range of field strength and frequency the droplets form rotating structures with a triangular order - rotating crystals. A model is developed to describe ensembles of several droplets, taking into account the hydrodynamic interactions between the rotating droplets in the presence of a solid wall below the rotating ensemble. A good agreement with the experimentally observed periodic dynamics for an ensemble of four droplets is obtained. During the rotation, the tips of the elongated magnetic droplets approach close to one another. An expression is derived that gives the magnetic interaction between such droplets by taking into account the coulombian forces between magnetic charges on the droplet tips.

## 1. Introduction

In recent years, systems based on out-of-equilibrium assemblies have gained considerable attention, see for example the recent review on field driven self-assemblies of colloids by active structuring [1].

In particular, the formation of spinning dynamical structures under rotating magnetic fields, leading to 2D rotating-crystals, have been studied in a series of works [2, 3, 4, 5] with different systems, showing several specificities of such "spinners". Ref. [2, 3] study a system of millimeter-sized magnetized disks, floating above a liquid-gas interface under the action of an inhomogeneous field (created by rotating permanent magnets). The repulsion between disks at finite Reynolds number results in a lift force between the rotating particles [2, 3]. The formation of rotating crystals of magnetic Janus colloids is described in [4]. There, a significant role is played by lubrication effects between rotating colloids. These effects are responsible for the dependence of the crystal angular velocity on the crystal size. In [5], an ensemble of magnetic droplets at the liquid-air interface is observed to form ordered dynamical structures under rotating field. Each droplet contains magnetic micron-sized particles which are chaining under magnetic field. The precessing field induces the rotation of the chains inside the droplets, which results in the global rotation of the droplets, which, in turn, auto-organize in a dynamical crystal.

Such 2D rotating crystals are also considered in the broader context of active matter. Let us mention here the rotating crystals formed by fast moving sulfur oxidizing bacteria Thiovolum Majus where the formation of the rotating crystals is due to the hydrodynamic and steric interactions between cells [6]. Moreover, a spontaneous breaking of the internal chirality leads to the development of rotational motion of self-propelling particles [7].

Here 2D rotating crystals of a new kind - ensembles of rotating needle-like magnetic droplets (micron-sized and re-sulting from a phase separation [8]) - are observed and described. The observation parameters are chosen in the region, where according to the sequence of bifurcations of such magnetic droplets, the system is above the transition oblate-prolate [8, 9]. For densely packed needle-like droplets in a rotating field, the axis ratios resulting from an interplay between the droplet breakup and coalescence have been examined by [10]. We present below the experimental observations of ensembles of such droplets under rotating field and their tentative modeling.

## 2. Experiment

### 2.1. Experimental setup

In the experiments we use a water-based ferrofluid made of maghemite nanoparticles (diameter $d = 7nm$, volume fraction $\varphi = 5\%$) that are stabilized with a citrate coating. A demixing in two coexisting phases can be induced in the ferrofluid by increasing its ionic strength, adding NaCl [11, 12]. In this way, droplets of highly concentrated phase are obtained coexisting with a surrounding medium at extremely low concentration of magnetic nanoparticles. These droplets exhibit under rather low static and dynamic fields, a series of shape instabilities as described in [8, 9].

We introduce $\approx 15 \mu l$ of the phase separated ferrofluid into a closed volume between two microscope slides separated by a double-sided tape of height $H \approx 130 \mu m$, and we observe it with a Leica DMI3000B microscope that is supplied with a home built coil setup. Droplets are filmed with a Basler acA1920-155um camera attached to the microscope, and the images are recorded in 8-bit gray-scale format. We track the droplet shape and position using a custom written program in MATLAB.

### 2.2. Determination of the magnetic droplet parameters

As in [9], we measure the properties of a few droplets (typically 10 microns in diameter) by elongating them along a magnetic field (of a few tens of gauss) and assuming an

---

axisymmetric ellipsoidal shape fitted with the theoretical relation from [13, 14]. From this we find that the relative magnetic permeability of the magnetic droplets is $\mu = 34 \pm 2$ and the surface tension is $\gamma = (8.2 \pm 0.4) \cdot 10^{-4} erg/cm^2$.

To determine their viscosity, the elongated droplets are let to relax in zero field, under their surface tension. The exponential relaxation of the droplet deformation parameter $D = (a-b)/(a+b)$ ($a$ and $b$ are the semi-axes of the droplet) is fitted as a function of time using the expression from [15] assuming small deformations $D \ll 1$ and an axisymmetric ellipsoidal shape. We get that the ratio of the droplet viscosity $\eta^i$ over the viscosity of the surrounding liquid $\eta^e$ is $\lambda = \eta^i/\eta^e = 10.1 \pm 2.5$.

## 2.3. Droplet behavior under a rotating magnetic field

When the phase separated ferrofluid sample is first placed under a microscope, some droplets are located close to the bottom of the cell due to their size and to their density being larger than that of the surrounding fluid, but with a distribution of heights. When a rotating magnetic field of magnitude $B = 60G$ and frequency $f = 15Hz$ is applied, we observe that the droplets arrange themselves in the same focal plane. By adjusting the focus of the microscope, we determined that the droplets are around $h = (9 \pm 1)\mu m$ above the bottom of the sample-cell.

In this applied rotating field, there is a critical volume $V_{crit} \approx \frac{\pi}{6}(3.5\mu m)^3$ for the droplets. If the volume of a particular droplet is larger than $V_{crit}$ the droplet flattens out into an oblate shape [8, 9] that sucks in and merges with smaller droplets in its immediate vicinity. However, if the droplet volume is below this critical value, they elongate and rotate with the same frequency as the field without coalescing. This is illustrated in Figure 1.

Once the at first unsteady motion has settled, the small-enough elongated droplets arrange themselves into crystal like structures (Figure 2) rotating in the direction of the magnetic field, albeit more slowly. The droplets in these crystals experience some oscillations around their lattice points, however the crystal structure remains intact and it rotates like a solid body (except sometimes for the most outer layer, where some drift is observed).

In regions where the droplet concentration is larger, the droplets form 2D triangular crystals that are too large to noticeably rotate. The centers of the droplets are Delaunay triangulated to clearly see the lattice structure. Figure 3) illustrates that the 2D structure is polycrystalline with a grain boundary marked by an array of dislocations.

The droplet lengths are more similar within a stationary single crystal than they were at first. Figure 4 illustrates a mechanism by which the droplets could achieve this. The length (and presumably also the volume) of a small droplet situated in between the lattice points of a large crystal increases over time. It is hard to assess the change of the droplet's width since it is close to the microscope spatial resolution and its determination is sensitive to errors from an ellipsoidal fit during the image processing.

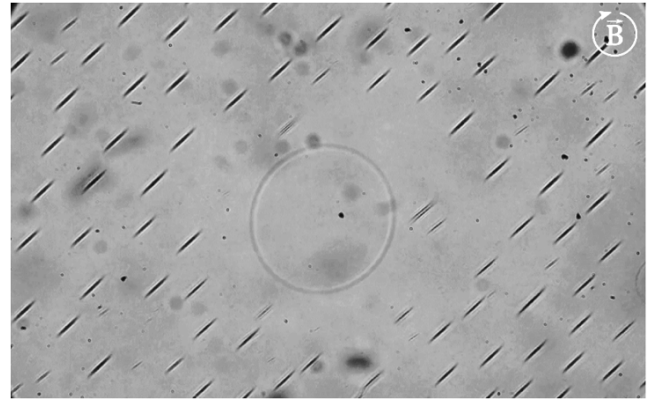

**Figure 1:** A view of droplets right after a rotating magnetic field is turned on. If a droplet's volume is below a critical value, it elongates and rotates with the field. If a droplet's volume is above a critical value, it flattens out into an oblate shape that also rotates with the field (seen at the center of the image). Field of view = $400 \times 250\mu m^2$. Magnetic field ($B = 60G$ and $f = 15Hz$) is rotating in the plane of observation

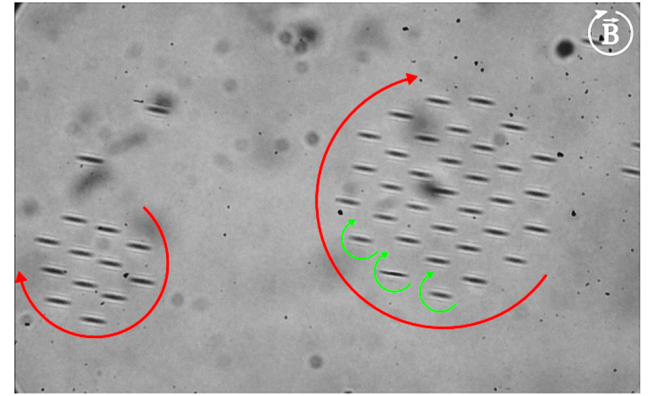

**Figure 2:** In time droplets organize themselves in 2D rotating crystals. Each individual droplet is rotating in the direction and with the frequency of the magnetic field. The crystal is also rotating in the direction of the magnetic field, but more slowly. Field of view = $400 \times 250\mu m^2$. Magnetic field ($B = 60G$ and $f = 15Hz$) is rotating in the plane of observation

## 3. Model

### 3.1. Description of the model

To understand the underlying interactions that lead to this ordering of the rotating elongated droplets, we develop a model below. The motion of the rotating droplet crystals is described by taking into account the flow each droplet produces. In the model, we assume that a droplet exerts a point torque at its center of mass producing a so called rotlet flow around it. Since the droplets are spinning close to the bottom of the sample, it is important to take into account the effects of the wall on the flow field.

We use the rotlet expression taking into account an infinite no-slip wall from [16]. In the case where the plane is located at $z = 0$ and a torque $\vec{\tau} = (0, 0, \tau)$ is applied at the point $(X, Y, Z)$, the velocity at $(x, y, z)$ simplifies and is

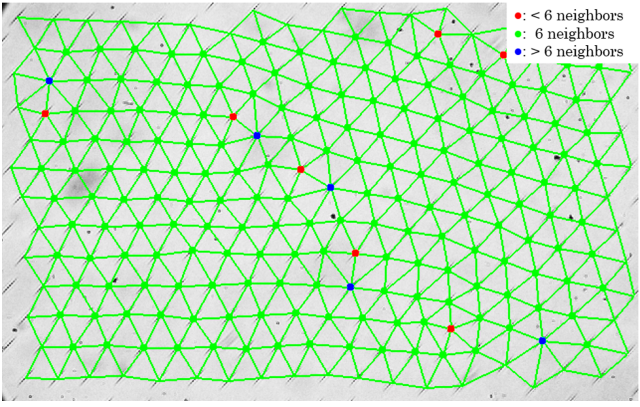

**Figure 3:** If the droplets are densely situated, they organize themselves in a large stationary 2D triangular lattice. Each droplet still rotates in the direction and with the frequency of the magnetic field, but the whole crystal is stationary. Field of view = $400 \times 250 \mu m^2$. Magnetic field ($B = 60 G$ and $f = 15 Hz$) and the droplets are rotating in the plane of observation

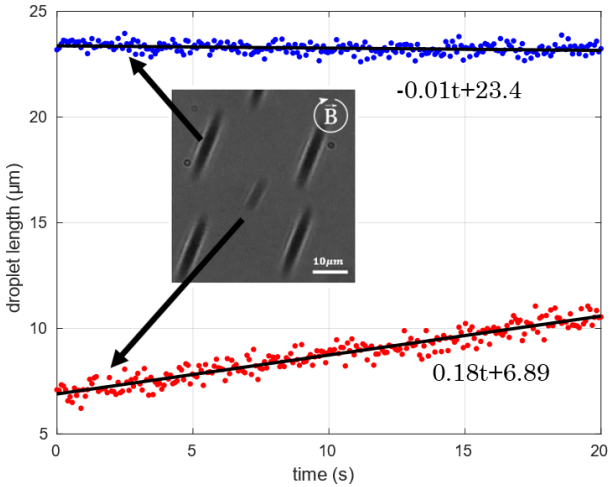

**Figure 4:** The lengths of two droplets within a large stationary crystal. The small droplet finds itself between bigger droplets that form the lattice. The length of the small droplet (red points) increases over time. The length of the large droplet (blue points) remains relatively static. The equations describe a linear fit.

given by

$$\vec{u} = \frac{1}{8\pi\eta}\left(\frac{\vec{\tau}\times\vec{r}}{|\vec{r}|^3} - \frac{\vec{\tau}\times\vec{R}}{|\vec{R}|^3}\right), \tag{1}$$

where $\vec{r} = (x-X, y-Y, z-Z)$ and $\vec{R} = (x-X, y-Y, z+Z)$, the viscosity $\eta$ here equals $\eta^e$. The velocity of the j-th droplet is calculated as the sum of rotlet fields produced by all other droplets at the location of the center of mass of j-th droplet, i.e. $\vec{v}_j = \sum_{i\neq j}\vec{u}_i$.

To determine the torque a rotating droplet exerts, we treat it as a rigid body. Even though the droplets are liquid and contract to spherical shapes if the magnetic field is removed, the approximation of a rigid particle is justified because the internal viscosity $\eta^i$ is 10 times larger than the viscosity $\eta^e$ of the surrounding medium. Since the droplets are well described by a slender body approximation (the aspect ratio is typically 18:1), and because they rotate with their rotational axis perpendicular to the infinite wall, the torque they produce is close to that of a slender body rotating in an infinite fluid (see Figure 15 in [17]). Therefore for simplicity, we use the expression given by [18] (page 56) for the torque of a rotating ellipsoid where the flow is in an infinite space.

Since this model does not take into account any radial forces of flows between two droplets, it does not impose an equilibrium distance between two droplets. Therefore, as an initial condition in the calculations, we place the droplets at positions taken from an experiment. We then let their positions evolve according to this model.

### 3.2. Tests of the model

To test the model, let us first look at the case of two droplets orbiting around a common center of mass. Experimentally the droplets are rotating with an angular velocity $\omega_{rot}^{exp} = (0.66 \pm 0.04)rad/s$. Using in the model the droplet length and starting with coordinates from the experimental images (as said above), we get a value for the angular velocity $\omega_{rot}^{theor} = (0.6 \pm 0.1)rad/s$ which is in good agreement with the experiment. We calculate the angular velocity of the rotating crystal by taking the average between all droplets of

$$\omega_{rot} = \frac{|\vec{r}\times\vec{v}|}{|\vec{r}|^2}, \tag{2}$$

where $\vec{r}$ is the vector from the center of mass of the crystal to that of the droplet and $\vec{v}$ is the droplet's velocity. The error for the model value is determined by varying input parameters within measurement error boundaries.

As shown in Figure 5 four magnetic droplets form a quadrilateral that rotates in the direction of the field and additionally its diagonal lengths oscillate. Again using droplet parameters from the experiment, we calculate the motion of the droplets and see that the rotation of the quadrilateral and the oscillation of the diagonal lengths is nicely reproduced as shown in Figure 6. The angular oscillation frequency of the experiment is found to be $\omega_{diag}^{exp} = (0.95 \pm 0.02)rad/s$ which coincides with the value from the model $\omega_{diag}^{theor} = (1.1 \pm 0.2)rad/s$. The experimental angular velocity of the crystal $\omega_{rot}^{exp} = (0.47 \pm 0.04)rad/s$ just barely coincides with the model value $\omega_{rot}^{theor} = (0.60 \pm 0.09)rad/s$. However, the ratio of the two cannot be described with the current model: $\omega_{diag}^{exp}/\omega_{rot}^{exp} = 2.0 \pm 0.2$ and $\omega_{diag}^{theor}/\omega_{rot}^{theor} = 1.71 \pm 0.02$. Interestingly, in the model, while the uncertainty in the droplet length is the primary source of error both for $\omega_{diag}$ and $\omega_{rot}$, the error in their ratio comes mostly from the uncertainty of the droplet position above the horizontal wall.

When calculating the motion of a large rotating crystal of droplets, we soon find that because there is nothing that forces an equilibrium distance between the droplets, the 2D crystal does not retain its regular shape and quickly disintegrates.

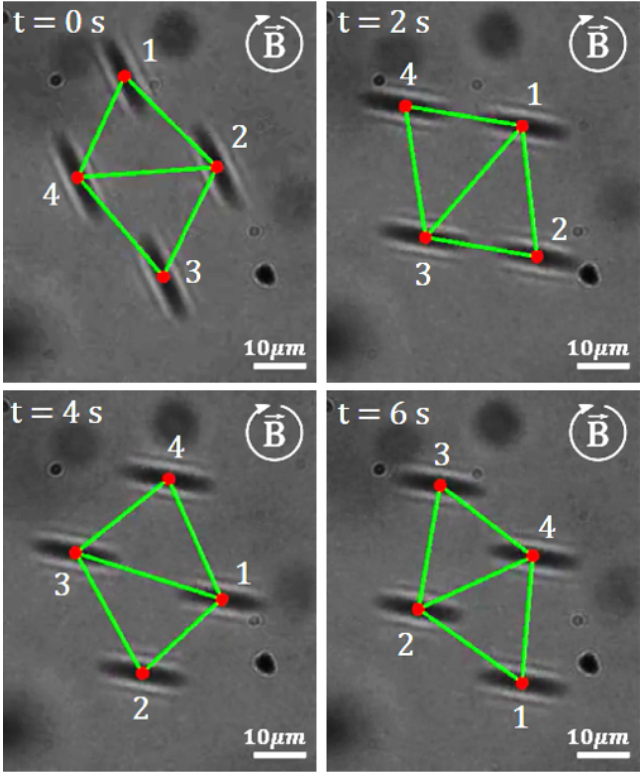

**Figure 5:** Four droplets form a quadrilateral rotating crystal with oscillating diagonal lengths. The droplets are labeled 1-4, and are Delaunay triangulated to better illustrate their motion. Snapshots are taken with 2 second increments. Magnetic field ($B = 60 G$ and $f = 15 Hz$) is rotating in the plane of observation

### 3.3. Suggestions for further improvement of the model

We propose that the equilibrium distance between the droplets is given by a balance between a magnetic attractive force and a radially outward secondary flow, as described in [19] for a rotating spherical particle. Some further work is still needed to write the secondary flow expression for a rotating elongated droplet near a wall. We do, however, propose an expression for the magnetic attraction (in Gaussian units).

Since the tips of two rotating elongated magnetic droplets with opposite magnetic charges $\pm \pi b^2 M$ ($M$ is the magnetization of the droplet) may come rather close to each other, the magnetic interaction between the droplets is given by a coulumbian interaction of the charges on their tips with coordinates $\vec{r}_{1,2} = (\pm a \cos(\Omega t), \pm a \sin(\Omega t))$; $\vec{r}_{3,4} = (r \pm a \cos(\Omega t), \pm a \sin(\Omega t))$ (the radius vector between centers of droplets is along $x$-axis).

Modeling the droplet as cylinder with cross-section radius $b$ and length $2a$, the energy of a coulumbian interaction

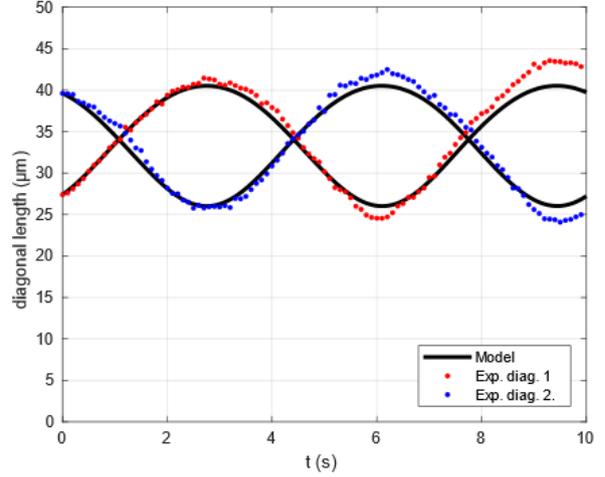

**Figure 6:** The diagonal lengths of the four droplet rotating crystal over time. The blue and red points are taken from experimental fits of the droplet positions. The black lines are the model results, which used the droplet position at $t = 0$ that were taken from the experiment and the droplet dimensions that were taken from the experiment within the margins of error such that the model best fits the experiment points.

$U$ may be written as $U = U_{13} + U_{24} + U_{14} + U_{23}$, where

$$U_{13} = U_{24} = \frac{(\pi b^2 M)^2}{r}$$

$$U_{14}(t) = -\frac{(\pi b^2 M)^2}{\sqrt{(r - 2a\cos(\Omega t))^2 + 4a^2 \sin^2(\Omega t)}} \quad (3)$$

$$U_{23}(t) = U_{14}(t + \pi/\Omega)$$

Taking the average with respect to the rotation period $2\pi/\Omega$ we obtain

$$U = \frac{2(\pi b^2 M)^2}{r}\left(1 - \frac{1}{2\pi}\int_0^{2\pi}\frac{dt}{\sqrt{(1 + 2a\cos(t)/r)^2 + 4a^2\sin^2(t)/r^2}}\right) \quad (4)$$

If $r >> 2a$, then we obtain $U = -\frac{m^2}{2r^3}$, where $m = 2a\pi b^2 M$ is the magnetic moment of the droplet. Thus at large distances between droplets, the interaction energy is given by the time averaged dipolar interaction energy.

The magnetic force between droplets is obtained by differentiation. It is shown in Figure 7, scaled by $m^2/8a^4$. Figure 7 shows that at close distance between droplets the dipolar interaction force is smaller by an order of magnitude than the force calculated using coulumbian interaction between effective magnetic charges. It is this strong attraction that balances the hydrodynamic flows in the formation of the 2D crystals of magnetic droplets.

## 4. Conclusions

Ensembles of magnetic microdroplets in a rotating magnetic field form structures with a regular triangular order -

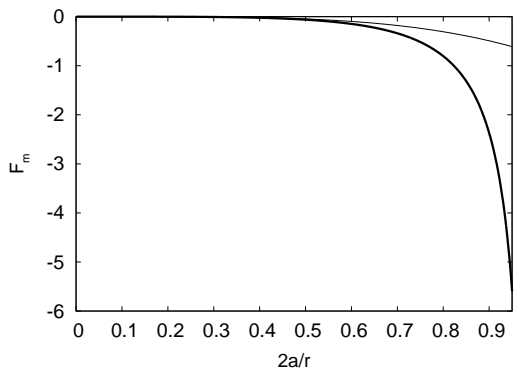

**Figure 7:** Dimensionless magnetic force as a function of the inverse distance between droplets. Exact relation - thick line, dipolar approximation - thin line.

rotating crystals. The distance between the droplets results from a balance between the magnetic attraction and the repulsion associated with the flow arising from inertial effects. Ensembles of several droplets (2,4) are well described by the model taking into account the hydrodynamic interactions between the droplets in the presence of a solid wall underneath. In large arrays, rows of dislocations are observed at the grain boundaries. It is shown that for an accurate description of the magnetic interaction between the droplets, the coulombian interaction between the magnetic charges on the tips of the elongated droplets must be considered.

## 5. Acknowledgments

We express gratitude to D. Talbot for the ferrofluid sample, and G. Kitenbergs for helpful experimental insights. The research was helped by the funding from the Embassy of France in Latvia, SIA "Mikrotīkls" and the French-Latvian bilateral program "Osmose" FluMaMi (n°40033SJ ; LV-FR/2019/5).